# NURE: An ERC project to study nuclear reactions for neutrinoless double beta decay

**Manuela Cavallaro**[1]
*INFN – Laboratori Nazionali del Sud*
*Via S. Sofia 62, 95125, Catania, Italy*
*E-mail:* manuela.cavallaro@lns.infn.it

**E. Aciksoz, L. Acosta, C. Agodi, N. Auerbach, J. Bellone, R. Bijker, S. Bianco, D. Bonanno, D. Bongiovanni, T. Borello, I. Boztosun, V. Branchina, M.P. Bussa, L. Busso, S. Calabrese, L. Calabretta, A. Calanna, D. Calvo, F. Cappuzzello, D. Carbone, E.R. Chávez Lomelí, M. Colonna, G. D'Agostino, N. Deshmuk, P.N. de Faria, C. Ferraresi, J.L. Ferreira, P. Finocchiaro, M. Fisichella, A. Foti, G. Gallo, U. Garcia, G. Giraudo, V. Greco, A. Hacisalihoglu, J. Kotila, F. Iazzi, R. Introzzi, G. Lanzalone, A. Lavagno, F. La Via, J.A. Lay, H. Lenske, R. Linares, G. Litrico, F. Longhitano, D. Lo Presti, J. Lubian, N. Medina, D. R. Mendes, A. Muoio, J. R. B. Oliveira, A. Pakou, L. Pandola, H. Petrascu, F. Pinna, F. Pirri, S. Reito, D. Rifuggiato, M.R.D. Rodrigues, A.D. Russo, G. Russo, G. Santagati, E. Santopinto, O. Sgouros, S.O. Solakcı, G. Souliotis, V. Soukeras, D. Torresi, S. Tudisco, R.I.M. Vsevolodovna, R. Wheadon, V.A.B. Zagatto**
*INFN – Laboratori Nazionali del Sud, Catania, Italy*
*INFN – Sezione di Catania, Italy*
*INFN – Sezione di Torino, Italy*
*INFN – Sezione di Genova, Italy*
*Dipartimento di Fisica e Astronomia, Università di Catania, Italy*
*Politecnico di Torino, Italy*
*Università degli Studi di Enna "Kore", Enna, Italy*
*CNR-IMM, Sezione di Catania, Italy*
*Universidade de Sao Paulo, Brazil*
*Universidade Federal Fluminense, Niteroi, Brazil*
*University of Ioannina, Ioannina, Greece*
*Universidad Nacional Autónoma de México, Mexico*
*University of Giessen, Germany*
*Akdeniz University, Antalya, Turkey*
*School of Physics and Astronomy, Tel Aviv University, Israel*
*IFIN-HH, Bucharest, Romania*

---

[1]Speaker

Neutrinoless double beta decay (0νββ) is considered the best potential resource to determine the absolute neutrino mass scale. Moreover, if observed, it will signal that the total lepton number is not conserved and neutrinos are their own anti-particles. Presently, this physics case is one of the most important research "beyond Standard Model" and might guide the way towards a Grand Unified Theory of fundamental interactions.

Since the ββ decay process involves nuclei, its analysis necessarily implies nuclear structure issues. The 0νββ decay rate can be expressed as a product of independent factors: the phase-space factors, the nuclear matrix elements (NME) and a function of the masses of the neutrino species. Thus the knowledge of the NME can give information on the neutrino mass scale, if the 0νββ decay rate is measured.

In the NURE project, supported by a Starting Grant of the European Research Council, nuclear reactions of double charge-exchange (DCE) will be used as a tool to extract information on the ββ NME. In DCE reactions and ββ decay, the initial and final nuclear states are the same and the transition operators have similar structure. Thus the measurement of the DCE absolute cross-sections can give crucial information on ββ matrix elements.

Introduction

Double charge-exchange reactions (DCE) are processes characterized by the transfer of two units of the isospin component (two protons transformed into two neutrons or vice versa), leaving the mass number unchanged. The initial and final nuclear states involved in DCE reaction and ββ decay are the same and the transfer operators have similar spin-isospin mathematical structure. Namely they both contain a Fermi, a Gamow-Teller and a rank-two tensor term. A relevant amount of linear momentum (of the order of 100 MeV/c) is available in the virtual intermediate channel in both processes. This is a crucial similarity since the nuclear matrix elements strongly depend on the momentum transfer and other processes (single charge-exchange reactions, 2νββ decay etc.) cannot probe this feature. Thus, even if the two processes are mediated by different interactions, the involved nuclear matrix elements could be connected and the determination of the DCE reaction cross-sections could give important information on the ββ matrix elements.

One should remind that a proportionality relation is well established at a level of few percent between single β decay strengths and single charge-exchange reaction cross-sections, under specific dynamical conditions. Indeed, single charge-exchange reactions are routinely used as a tool to determine Fermi and Gamow-Teller transition strengths for single β decay, as demonstrated by several works [1-7]. However, studying the link between ββ-decay strengths and DCE cross-sections is a not trivial task and requires a strong effort.

Experimental attempts were done in the past to perform DCE reactions [8], [9]. However, most of them were not conclusive because of the very poor yields in the measured energy spectra and the lack of angular distributions, due to the very low cross-sections involved. High resolution spectra and angular distributions are in fact crucial to identify the transitions of interest and separate the direct reaction mechanism among other possible multi-step processes that can hide the desired information [10], [11].

The advent of new facilities based by large acceptance and high resolution spectrometers have allowed to obtain very promising results in this field. One should mention for example the recent measurements at RIKEN [12], [13] and the results of the experiment at INFN-LNS on $^{40}$Ca($^{18}$O,$^{18}$Ne)$^{40}$Ar [14], [15]. In particular, the experiment at INFN-LNS has shown that high resolution energy spectra and angular distribution can be accessed and matrix elements extracted within a promising level of accuracy at least for the $^{40}$Ca case.

**1 The idea**

NURE plans to carry out a campaign of experiments using accelerated beams on different target candidates for 0νββ decay in the 5-years duration of the project. The DCE channel will be populated using ($^{18}$O,$^{18}$Ne) and ($^{20}$Ne,$^{20}$O) reactions by the MAGNEX large acceptance spectrometer, which is designed to measure very suppressed reaction channels at high resolution and at zero degree [16]. The complete net involving the single charge-exchange and multi-step transfers characterized by the same initial and final nuclei will be also measured to study the reaction mechanism. The absolute cross-sections will be extracted.
The Superconducting Cyclotron accelerator at INFN-LNS provides all the required beams for the proposed project, namely $^{18}$O and $^{20}$Ne at energies ranging from 10 MeV/u to 60 MeV/u with an excellent energy resolution and emittance.

MAGNEX allows the identification of heavy ions with high mass (1/200), angle (0.2°) and energy (1/1000) resolutions, within a large solid angle (50 msr) and momentum range (25%). It also allows to measure at zero degrees, which is the most important region for spectroscopic studies, thus creating ideal laboratory conditions. High resolution measurements of other reactions characterized by cross-sections falling down to tens of nb/sr were already performed by this setup [17], [18]. The crucial issue of MAGNEX is the implementation of the powerful technique of trajectory reconstruction, which allows to solve the equation of motion of each detected particle to high order (10$^{th}$ order). In this way an effective compensation of the high order aberrations induced by the large aperture of the magnetic elements is achieved. The use of the sophisticated data reduction approaches based on the differential algebra is a unique feature of MAGNEX, developed in the past years [19-22]. This guarantees the above mentioned performances and its relevance in the research of heavy-ion physics [23], [24].

Two pairs of target nuclei of interest as candidates for $0\nu\beta\beta$ will be studied in both $\beta^-\beta^-$ and $\beta^+\beta^+$ direction: the $^{76}$Ge-$^{76}$Se and the $^{116}$Cd-$^{116}$Sn pairs. These targets represent hot cases in the research for $0\nu\beta\beta$ decay and belong to two different classes of nuclei: those in which protons and neutrons occupy the same major shells ($^{76}$Ge-$^{76}$Se) and those in which they occupy different major shells ($^{116}$Cd-$^{116}$Sn). Moreover, the magnitude of the Fermi matrix element, which is related to the overlap of the proton and neutron wave functions, is different in these two classes of nuclei, being large in the former and small in the latter case. The study of the different behaviour of the nuclear matrix elements in these systems is interesting. These nuclei have been chosen because the transition to the first excited states can be well separated from the ground states by the MAGNEX resolution (being the first excited states at 562 keV for $^{76}$Ge, 559 keV for $^{76}$Se, 1.29 MeV for $^{116}$Sn and 513 keV for $^{116}$Cd) and also because the production technologies of these targets are already available.

## 2 A broader view: the NUMEN program

The possibility to extract data-driven information on $0\nu\beta\beta$ nuclear matrix elements is an ambitious line of research which has been conceived and is developing in a broader context, characterized by a longer time scale and a larger collaboration: the NUMEN program [25].

In the present experimental conditions, due to the limitation arising from the tiny cross-sections of the processes of interest, only very few systems can be measured within the 5-years project. In order to systematically explore all the nuclei candidates for $0\nu\beta\beta$, a beam intensity at least two orders of magnitude higher than the present must be achieved. As a consequence, major upgrades of the detector technologies (3D ion tracker, particle-identification wall, gamma-ray array, …) must be developed [26], [27]. Also the target technology must be upgraded, to avoid the damage of the thin films due to the high temperature involved. The front-end and readout electronics must take into account the high number of channels and the expected rate of detected events. Moreover, a deep and complete investigation of the theoretical aspects connecting nuclear reaction mechanisms and nuclear matrix elements must be carried out.

All of these issues and other aspects related to the high intensity beam physics are the crucial branches of the NUMEN program.

**Acknowledges**

This project has received funding from the European Research Council (ERC) under the European Union's Horizon 2020 research and innovation programme (grant agreement No 714625)